\documentclass[%
 reprint,
superscriptaddress,
 amsmath,amssymb,
 aps,
pra,
longbibliography
]{revtex4-1}
\usepackage[utf8]{inputenc}
\usepackage{graphicx}
\usepackage{dcolumn}
\usepackage{bm}
\usepackage{hyperref}
\usepackage{amsmath}
\usepackage{amssymb}
\usepackage{pdfpages}
\usepackage{float}
\usepackage{braket}
\usepackage{tikz}
\makeatletter
\AtBeginDocument{\let\LS@rot\@undefined}
\makeatother
\newcommand{\kl}[1]{\ensuremath{ \left( #1 \right) }}

\begin{document}
\title{Topologically non-trivial valley states in bilayer graphene quantum point contacts }

\author{Hiske Overweg}
\email{overwegh@phys.ethz.ch}
\affiliation{%
Solid State Physics Laboratory, ETH Zürich,~CH-8093~Zürich, Switzerland}
\author{Angelika Knothe}
\affiliation{%
National Graphene Institute, University of Manchester,
Manchester M13 9PL, UK}
\author{Thomas Fabian}
\author{Lukas Linhart}
\affiliation{Institute for Theoretical Physics, Vienna University of Technology, A-1040 Vienna, Austria}

\author{Peter Rickhaus}
\author{Lucien Wernli}
\affiliation{%
Solid State Physics Laboratory, ETH Zürich,~CH-8093~Zürich, Switzerland}

\author{Kenji Watanabe}
\author{Takashi Taniguchi}
\affiliation{National Institute for Material Science, 1-1 Namiki, Tsukuba 305-0044, Japan}
\author{David Sánchez}
\affiliation{Institute for Cross-Disciplinary Physics and Complex Systems IFISC  (UIB-CSIC),
07122 Palma de Mallorca, Spain}
\author{Joachim Burgdörfer}
\author{Florian Libisch}
\affiliation{Institute for Theoretical Physics, Vienna University of Technology, A-1040 Vienna, Austria}
\author{Vladimir I. Fal'ko}
\affiliation{%
National Graphene Institute, University of Manchester,
Manchester M13 9PL, UK}
\author{Klaus Ensslin}
\author{Thomas Ihn}
\affiliation{%
Solid State Physics Laboratory, ETH Zürich,~CH-8093~Zürich, Switzerland}
 \email{overwegh@phys.ethz.ch}

\date{\today}

\begin{abstract}
We present measurements of quantized conductance in electrostatically
induced quantum point contacts in bilayer graphene. The application of
a perpendicular magnetic field leads to an intricate pattern of lifted
and restored degeneracies with increasing field: at zero magnetic
field the degeneracy of quantized one-dimensional subbands is four,
because of a twofold spin and a twofold valley degeneracy. By
switching on the magnetic field, the valley degeneracy is lifted.  Due
to the Berry curvature states from different valleys split linearly in
magnetic field.  In the quantum Hall regime fourfold degenerate
conductance plateaus reemerge. During the adiabatic transition to the quantum
Hall regime, levels from one valley shift by two in quantum number with respect to the
other valley, forming an interweaving pattern that can be
reproduced by numerical calculations.
\end{abstract}
\maketitle

Conductance quantization in one-dimensional channels is among the
cornerstones of mesoscopic quantum devices. It has been observed in a
large variety of material systems, such as $n$-type
GaAs~\cite{Wees1988,Wharam1988}, $p$-type
GaAs~\cite{Rokhinson2006,Danneau2006}, SiGe~\cite{Tobben1995},
GaN~\cite{Chou2005a}, InSb~\cite{Goel2005}, AlAs~\cite{Gunawan2006}
and Ge~\cite{Mizokuchi2018}. Typically spin degeneracy leads to
quantization in multiples of $2~e^2/h$. In single and bilayer graphene
both steps of $2~e^2/h$ and $4~e^2/h$ have been reported
\cite{Tombros2011,Terres2016a,Kim2016a,Allen2012,Goossens2012a,Overweg2017},
although a fourfold degeneracy is expected due to the additional
valley degree of freedom.  Here we present data for three quantum
point contacts (QPCs) which display (approximately) fourfold degenerate
modes both at zero magnetic field and in the quantum Hall regime, and
twofold degenerate modes in the transition region.  The Berry
curvature in gapped bilayer graphene induces an orbital
magnetic moment for the states selected by the quantum point
contact. The valleys therefore split linearly in a weak magnetic field
and conductance steps of $2 e^2/h$ emerge. The adiabatic evolution of
conduction steps to the quantum Hall regime reveals a universal level
crossing pattern: state energies in one valley shift by two with
respect to those of the other valley due to the chirality of the
effective low-energy Hamiltonian in the $K_+$ and $K_-$ valley, a
general feature of Dirac particles in even spatial
dimensions~\cite{Redlich1984}. Related topological effects involving
the valley degree of freedom have recently been discussed in bilayer \cite{Sui2015,Cortijo2012,Novoselov2006,Rao2013} and trilayer graphene~\cite{Taychatanapat11} . The lifting and restoring of
level degeneracies is explained in detail by two complementary
theoretical models. These results are the basis for a detailed
understanding of conductance quantization and tunneling barriers in bilayer graphene, enabling
high-quality quantum devices.

%
\begin{figure}
\centering
\includegraphics[width=0.5\textwidth]{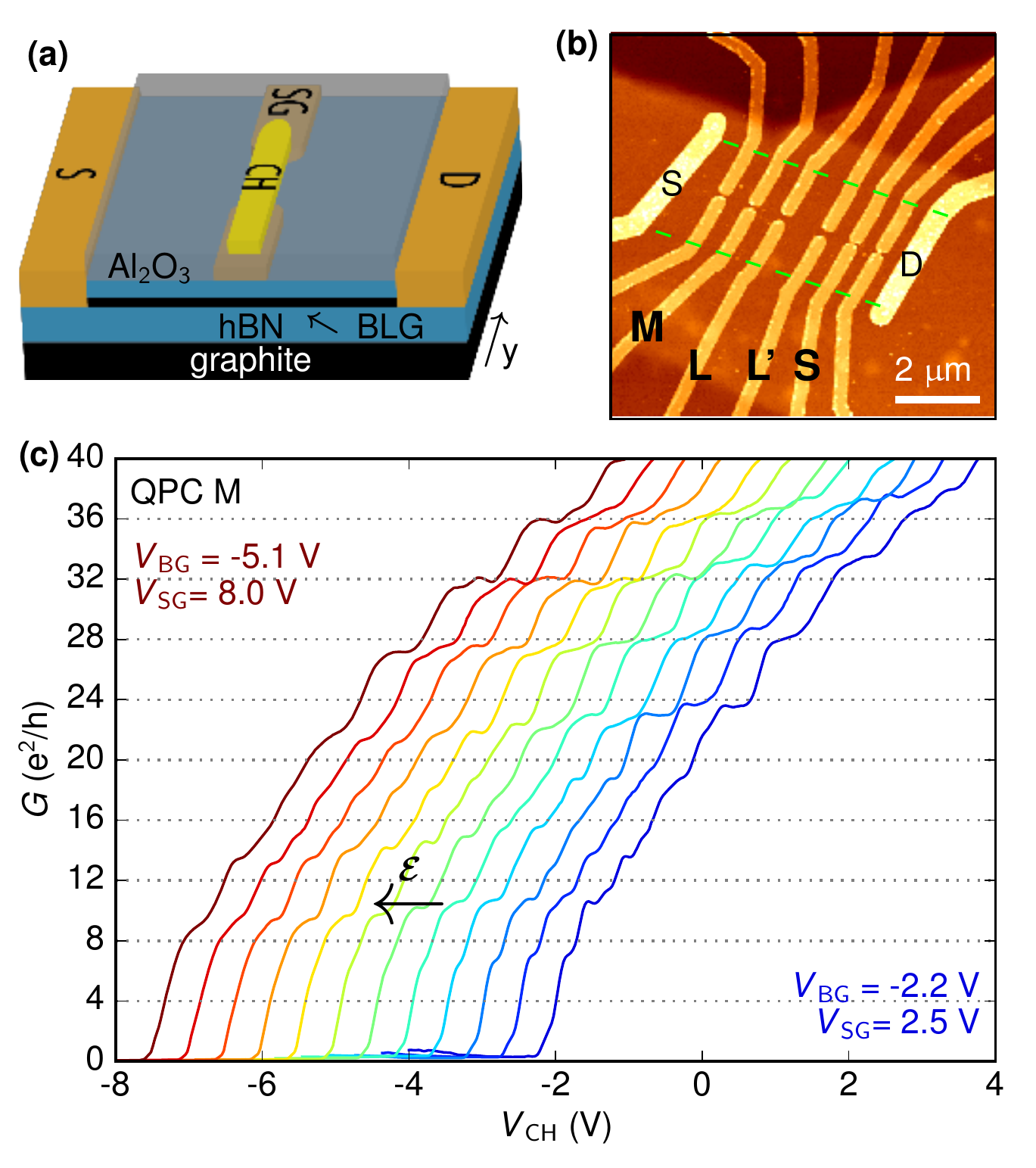}
\caption{(a) Schematic of the device consisting of bilayer graphene
  encapsulated in hBN on top of a graphite back gate. On top of the
  device split gates were evaporated. A layer of Al$_2$O$_3$ serves to
  separate the split gates from the channel gate. (b) Atomic force microscopy image of the device. Green dashed
  lines denote the edges of the bilayer graphene flake. Contacts are
  labeled S and D. Six pairs of split gates are situated between the
  contacts. (c) Conductance as a
  function of channel gate voltage for various combinations of the
  split gate and back gate voltage, showing conductance plateaus with
  a step size of $\Delta G = 4 e^2/h$ for large quantum numbers.}
\label{fig:device}
\end{figure}


The device geometry is similar to the one employed in our demonstration of full
pinch-off of bilayer graphene quantum point contacts
\cite{Overweg2017}. A bilayer graphene (BLG) flake was encapsulated
between hexagonal boron nitride layers (hBN), using the van der Waals
pick-up technique \cite{Wang2013}, and deposited onto a graphite flake
(see Figure ~\ref{fig:device}a for a schematic of the final device
geometry). The graphene layer was contacted with Cr/Au contacts and a
top gate pattern, consisting of six pairs of split gates (SG) with
spacings ranging from 50 nm to 180 nm, was evaporated.  On top of the
device a layer of Al$_2$O$_3$ and finally the channel gates (CH) were
deposited. An atomic force microscopy image of the sample, recorded
prior to the deposition of the channel gates, is shown in
Fig.~\ref{fig:device}b.  In the present manuscript, we show data from
three QPCs: QPC S (50 nm split gate separation), QPC M (80 nm) and QPC
L (180 nm).

By applying voltages of opposite sign to the graphite back gate and
the split gates, a band gap is
induced~\cite{McCann2006a} in the bilayer
graphene. In Ref.~\citenum{Overweg2017} we demonstrated that this
suppresses transport below the split gates. Hence a constriction is
formed, in which the charge carrier density can be tuned by the
channel gate voltage. During the measurements only a single pair of
split gates was biased at a time to form a QPC. The measurements were
performed at $ T = 1.7~$K.

The conductance of QPC M (80 nm wide) as a function of channel gate
voltage is shown in Fig.~\ref{fig:device}c for various combinations of
the split and back gate voltage. For each curve, a series resistance
was subtracted which corresponds to the resistance measured at the
same back gate voltage with uniform charge carrier density throughout
the sample. The traces show several plateaus with a typical step size
of $\Delta G = 4~e^2/h$, in particular for large quantum numbers, as
previously reported in Ref.~\citenum{Overweg2017}. Similar results
have been found for QPC L and L' (180 nm wide) with a smaller spacing
in gate voltage between the plateaus, in agreement with the wider
channel, and for QPC S (50 nm wide) with a larger spacing. For the
employed range of gate voltages, the displacement field $\mathcal{E}$
does not significantly change the observed plateau sequence. Below $G
= 24~e^2/h$ we observe several kinks which cannot be identified as
plateaus and some plateaus occurring below the expected conductance
values. Reduced screening of the disorder potential in this low
density regime might play a role. Simulations of the electrostatic
potential~\cite{Overweg2017} show that in this regime the confinement
potential is shallow. From a theoretical perspective the
non-monotonicity of the dispersion relation, which becomes more
pronounced for larger gaps and wider channels, can lead to additional
degeneracies for low mode numbers, possibly explaining the absence of
a plateau at $G = 4~e^2/h$~\cite{Knothe2018}.

\begin{figure*}
\centering
\includegraphics[width=1\textwidth]{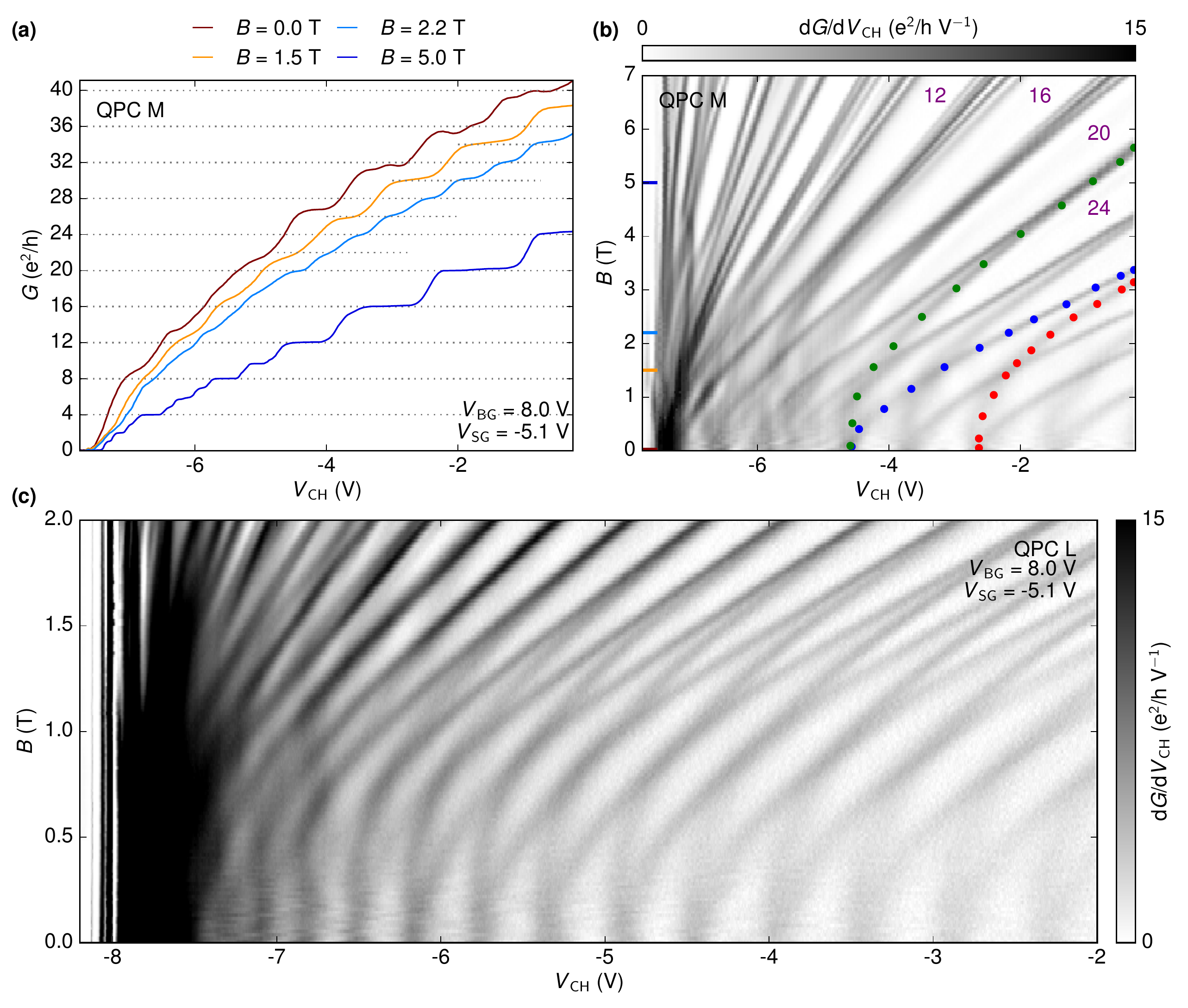}
\caption{(a) Conductance of QPC M as a function of $V_\mathrm{CH}$ for
  various magnetic field strengths. Several quantization sequences are
  observed. (b) Transconductance of QPC M as a function
  $V_\mathrm{CH}$ and magnetic field. A pattern of mode splittings
  (see green and blue dotted modes) is observed. Numbers in purple
  indicate the conductance values in the quantum Hall regime. (c)
  Transconductance of QPC L as a function $V_\mathrm{CH}$ and magnetic
  field. A similar pattern of mode crossings is observed in a smaller
  magnetic field range than for QPC M.}
\label{fig:fans}
\end{figure*}
 
The conductance of QPC M as a function of channel gate voltage for
several magnetic field strengths (Figure~\ref{fig:fans}a) features a
plateau sequence at $B =~0~T$ described by $G = 4 N e^2/h$ with
integer $N$. Increasing the magnetic field to a value of $B = 2.2~$T
changes the plateau sequence to $G = 2 N e^2/h$. At $B = 5~$T the
conventional sequence of Landau levels of BLG is observed, with $G = 4
N e^2/h$. In the lowest two Landau levels a lifting of the fourfold
degeneracy can be observed.  Around $B=1.5~$T, during the transition
to the Hall regime, the fourfold degeneracy is restored: the sequence
is shifted to $G = (4 N + 2) e^2/h$. This is most clearly visible for
the modes for which $G \geq 22~e^2/h$.

To further investigate this transition we inspect the transconductance
as a function of channel gate voltage and magnetic field (see
Fig.~\ref{fig:fans}b). Mode transitions show up as dark lines, which
start out vertically in low magnetic fields, but bend toward more
positive gate voltages above $B = 1~$T. This phenomenon, known as
magnetic depopulation and observed for instance in high quality GaAs,
is due to the transition from electrostatic confinement to magnetic
confinement. What is unusual however, is the pattern of mode
splittings and mode crossings. 

The same pattern can be observed for the wider QPCs
(Fig.~\ref{fig:fans}c), where the fourfold degeneracy is already
restored at 2~T because of the wider channel. Although the lowest
modes are hard to resolve, a robust pattern of mode crossings can be
observed for many higher modes.  Similar patterns could be observed
for various displacement fields inside the channel and also for a
$p$-doped channel (see Appendices~\ref{sec:displ} and
\ref{sec:ptype}).

\begin{figure*}
\centering
\includegraphics[width=0.8\textwidth]{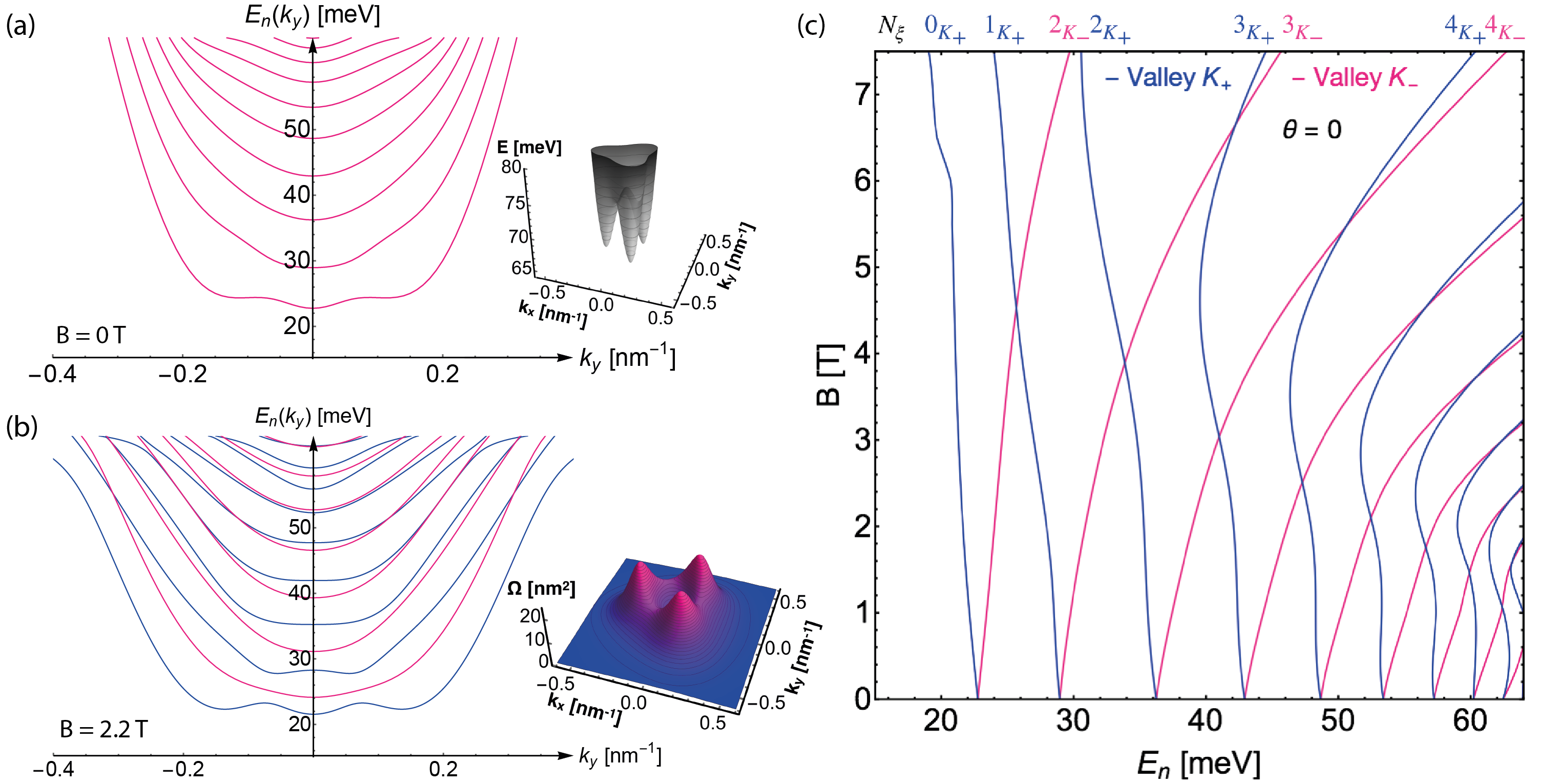}
\caption{(a) Bandstructure (conduction band) at $B=0$ T of BLG in the presence of a confinement potential $U(x)$ and a modulated gap $\Delta(x)$ as described in the text showing a discrete, valley degenerate mode spectrum. Inset: lowest conduction band of homogeneous gapped BLG (K$_-$ valley) for $\Delta_0=150$ meV with three minivalleys forming around the $K$-point.
(b) Bandstructure (conduction band)  of the channel at $B=2.2$ T, where symmetry between valleys is broken. The valley splitting at small magnetic fields is proportional to the magnetic field. Inset: Berry curvature $\Omega$  of the corresponding states with non-zero peaks in the three minivalleys. 
(c) Magnetic field dependence of the subband edges of the conduction bands in the electron channel. The nontrivial Berry curvature of the zero-field states implies a non-zero orbital magnetic moment $M$ of the states, $M\propto\Omega$, which induces the linear in magnetic field splitting at small magnetic fields. At high magnetic fields the levels evolve into the LLs of gapped BLG.}
\label{fig:man}
\end{figure*}

\begin{figure}
\centering
\includegraphics[width=\columnwidth]{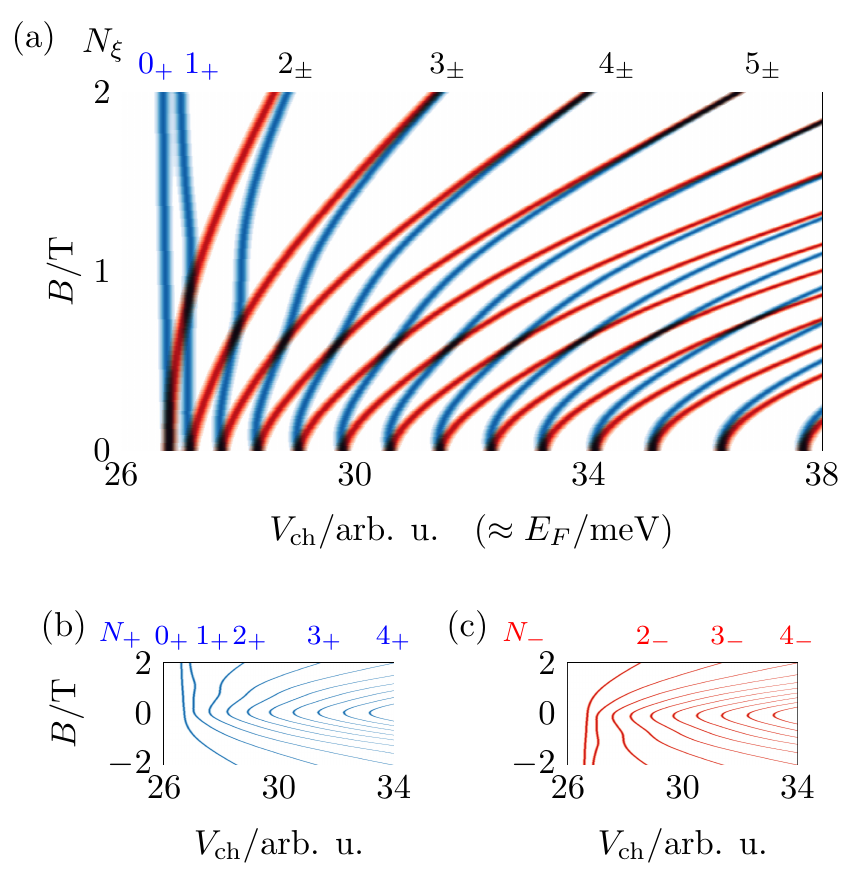}
\caption{a) Differential conductance $\mathrm{d}G/\mathrm{d}E$ of a
  180 nm wide BLG nanoribbon, including a thermal
  smoothening of 1.7 K. b) and c) show separately the contributions
  from the two valleys $K_+$ and $K_-$ at low energies. 
  The channel voltage is determined through the relation $V_{ch} \propto \sqrt{E_F^2 - (\Delta/2)^2}$, with $\Delta/2 =$~25~meV.}
\label{f:diffconductance}
\end{figure}

To elucidate the evolution of the conductance steps with magnetic
field, we simulate the experimental setup using two independent, complementary theoretical approaches, $k
\cdot p$ theory~\cite{McCann2006a} and tight-binding calculations~\cite{Libisch12} (see the
supplement for technical details). Both approaches agree well with
each other and the experiment, highlighting the robustness of the
observed features and the validity of our two modelling approaches. They reproduce and explain the observed  low-field splitting (Fig.~\ref{fig:man}) and the level
crossing pattern (Fig.~\ref{f:diffconductance}).

Here, we use soft
electrostatic confinement provided by a transverse electric field both at $B=0$ and at a finite magnetic field.
The obtained magnetic field dependence of the miniband edges
represents the closest spectral analogue of the experimentally
measured transconductance spectrum. We chose the potential landscape for $k
\cdot p$ theory by matching the mode spacing extracted
from finite bias measurements of QPC M at $B=0~$T (see
Appendix~\ref{sec:fb}).
Note that in the experiment the channel gate voltage influences not only the Fermi level, but also the shape of the confinement potential and the size of the displacement field inside the channel. To obtain one to one agreement between the calculation and the experimental results, a self-consistent potential would be required.

 At zero magnetic field, we find spin- and
valley-degenerate spectra (Fig.~\ref{fig:man}a) in agreement with the
experimentally observed step size of $\Delta G = 4~e^2/h$
(Fig.~\ref{fig:fans}a).  The subband edges (for small mode numbers)
are situated at finite momenta, reminiscent of the three mini-valleys
in gapped BLG in the presence of trigonal
warping~\cite{McCann2006a,Varlet2014a}. When switching on a magnetic field, the interlayer asymmetry leads
to  valley splitting of electron subbands, clearly seen in the band
structure computed for $B = 2.2~$T (Fig.~\ref{fig:man}b, blue and
magenta subbands). This lifting of   valley degeneracy is in
agreement with the measured step size of $\Delta G = 2~e^2/h$ (see
Fig.~\ref{fig:fans}a). The linearity of the valley splitting of the subband edges with $B$, Fig.~\ref{fig:man}b, is related to the fact that the zero-field states $\ket{n _\xi}$ (with transverse quantum number $n =0,1,\ldots$ in the $K_\xi$ ($\xi = \pm$) valley) of trigonally warped gapped BLG~\cite{McCann2006a, Varlet2014a} carry
non-trivial Berry curvature (see insets of Fig.~\ref{fig:man} and in Appendix~\ref{sec:bulk}) and, consequently, a finite magnetic
moment, $M_z$. For larger displacement fields,
the Berry curvature becomes more spread out in $k$-space around the
K-points, affecting several of the lowest modes.

In the high magnetic field regime, where the magnetic
length is smaller than the channel width, the subbands in the channel become drifting states in the BLG  Landau
levels (LLs) $\ket{N_{\xi}}$, where $N$ now indicates the LL index.  The LL spectrum of BLG has a pair of special states $N=0,1$, that appear at zero energy in ungapped BLG with the wave functions residing on different layers in the opposite valleys. After the displacement field introduces a layer asymmetry gap, these states split apart by $\Delta$, resulting in the two lowest conduction band subbands belonging to only one valley, e.g. $K_+$ (then, the highest valence band subbands would be from valley $K_{-}$). The other LLs in both valleys with $N\ge2$ have approximately the same weight on the sublattices in the two layers and very close energies. Such an asymptotic behavior corresponds to the evolution of the subbands such that subbands $(n+2)_{K_+}$  eventually merge with subbands $n_{K_-}$ upon an increase in magnetic field as shown in Fig.\ref{f:diffconductance}a. For $B < 0$, the same pattern emerges with the two valleys interchanged (see Fig.~\ref{f:diffconductance}b,c).

Note that the absence of hard edges characteristic for the present
electrostatically defined bilayer constriction is critical for
observing the interweaving pattern of crossing states.
In rough-edged constrictions broken valley symmetry due to scattering 
quickly obscure the underlying pattern. 
These difficulties aside, a similar crossing pattern appears in principle in single layer graphene,
as we have verified numerically for an ideal constriction (see Appendix \ref{app:mono}).

In conclusion, we reported on the experimental observation of the mode crossing pattern during the evolution from size quantization to the Hall regime in BLG QPC.
A valley splitting linear in magnetic field could be explained by a non-trivial
orbital magnetic moment of states in gapped BLG. Our
experimental results could be reproduced by numerical simulations.

\clearpage
\section*{Acknowledgements}
We thank A.~Rebhan for fruitful discussions.
We acknowledge financial support from the European Graphene Flagship, the Swiss National Science Foundation via NCCR Quantum Science and Technology, ERC Synergy Hetero 2D, WWTF project MA14-002 and MECD. Calculations
were performed on the Vienna Scientific Cluster
(VSC). Growth of hexagonal boron nitride crystals was supported by the
Elemental Strategy Initiative conducted by MEXT, Japan and the CREST
(JPMJCR15F3), JST.

\bibliography{PhD-depop}
\clearpage
\appendix

\section{Model Hamiltonian}
\label{sec:model}
The four-band model Hamiltonian of BLG (BLG) is given by~\cite{McCann2006a, Varlet2014a}
\begin{align}
\nonumber &H^{\xi}_{BLG}=\\
\xi
&\setlength{\arraycolsep}{-5pt} \begin{pmatrix} 
 \xi U(x)-\frac{1}{2}\Delta(x) & v_3\pi & 0 &v \pi^{\dagger}\\
 v_3 \pi^{\dagger}& \xi U(x)+\frac{1}{2}\Delta(x) & v\pi &0\\
 0 & v\pi^{\dagger} & \xi U(x)+\frac{1}{2}\Delta(x) & \xi \gamma_1\\
 v\pi & 0 & \xi \gamma_1 & \xi U(x)-\frac{1}{2}\Delta(x)
\end{pmatrix},
\label{eqn:H}
\end{align}

written in the basis $\Phi_{K^+}=(\Psi_{A1},\Psi_{B2},\Psi_{A2},\Psi_{B1})$ or $\Phi_{K^-}=(\Psi_{B2},\Psi_{A1},\Psi_{B1},\Psi_{A2})$ in the two valleys $K^+$ (for $\xi=+1$), and $K^-$ (for $\xi=-1$). The diagonal terms in this Hamiltonian account for the spatially modulated confinement potential $U(x)$ and the modulated gap $\Delta(x)$:
\begin{equation}
U(x)=\frac{U_0}{\cosh{\frac{x}{L}}}, \hskip20pt \Delta(x)=\Delta_0-\beta\frac{\Delta_0}{\cosh{\frac{x}{L}}},
\label{eqn:UD}
\end{equation}
where we chose $U_0=-30$ meV, $L=20$ nm, and $\beta=0.3$, in accordance with the parameters of the experimental probes. Furthermore, $\pi=p_x+ip_y$, $\pi^{\dagger}=p_x-ip_y$, with $\mathbf{p}=-i\hbar\nabla-\frac{e}{c}\mathbf{A}$  and for the velocities and hoppings we use $v=1.0228*10^6$ m/s, $v_3=1.2299*10^5$ m/s, and  $\gamma_1=381$ meV~\cite{Varlet2014a}.

\section{Bulk properties}
\label{sec:bulk}
\begin{figure}
 \centering
\includegraphics[width=0.4\textwidth]{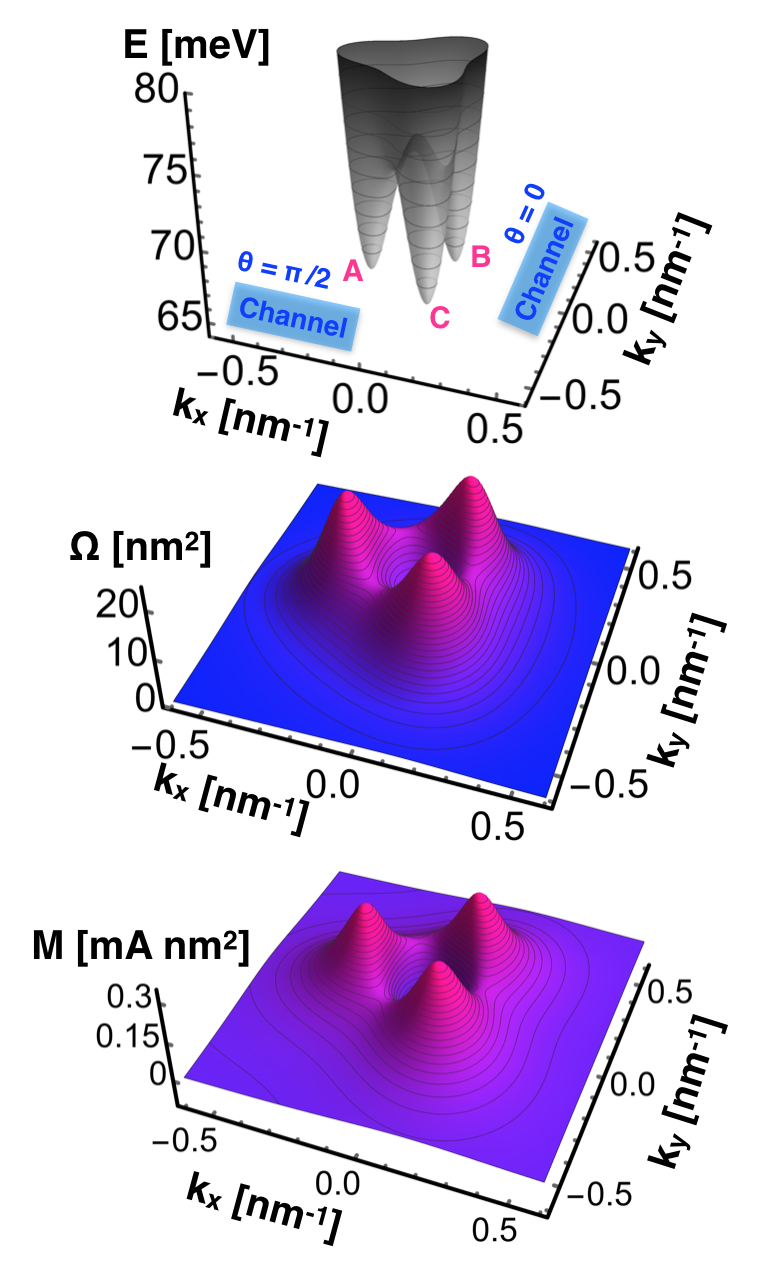}
 \caption{Top: Dispersion of homogeneous, gapped BLG in the $K_-$ valley (lower conduction band for $\Delta=150$ meV). The letters $A, B$, and $C$ label the three anisotropic minivalleys that form around each $K$-point due to trigonal warping. The blue stripes indicate the orientation of the channel at angles $\theta=0$ or $\theta=\pi/2$, respectively.  Bottom: Corresponding Berry curvature $\Omega$ and magnetic moment $M$ of the states of the lower conduction band in the $K_-$ valley for $\Delta=150$ meV. In the $K_+$ valley the sign of both $\Omega$ and $M$ is reversed. }
\label{fig:Berry}
\end{figure}

In the limit $x\to\infty$ the Hamiltonian of Eg.~\ref{eqn:H} describes the properties of homogeneous gapped BLG. The  dispersion is given by the four valley degenerate bands~\cite{McCann2006a}
\begin{align}
\nonumber E_{\pm}^{\alpha}&=\pm    \frac{\gamma_1^2}{2} +\frac{\Delta^2}{4} +(v^2+\frac{v_3^2}{2})\hbar^2k^2 \\
\nonumber&+ (-1)^{\alpha} \Bigg(  \frac{(\gamma_1^2-v_3^2\hbar^2k^2)^2}{4} +v^2\hbar^2k^2 [\gamma_1^2 +\Delta^2 +v_3^2\hbar^2k^2] \\
&+2\xi \gamma_1 v_3 v^2 \hbar^3k^3 \cos3\varphi     \Bigg)^{\frac{1}{2}},
\end{align}
where $\alpha=1,2$, and $\mathbf{k}=k(\cos\varphi,\sin\varphi)$. In the top panel of Fig.~\ref{fig:Berry} we plot the lower conduction band ($\alpha=1$) in the $K_-$ valley for a gap of $\Delta_0=150$ meV demonstrating the effect of trigonal warping induced by $v_3\neq0$ causing the dispersion to form three mini-valleys around each $K$-point.  The states of gapped BLG carry a non-trivial Berry curvature and orbital magnetic moment. From the Bloch functions of the $n$th band the magnitude of the corresponding Berry curvature $\mathbf{\Omega}_n(\mathbf{k})=\Omega_n \mathbf{e}_z$ and the orbital magnetic moment $\mathbf{M}_n(\mathbf{k})=M_n\mathbf{e}_z$ can be computed according to Refs.~\cite{Xiao2010,Chang1995} 
\begin{align}
\nonumber{\Omega}_n(\mathbf{k})&=i\langle\mathbf{\nabla}_{\mathbf{k}}\Phi_n|\times|\mathbf{\nabla}_{\mathbf{k}}\Phi_n\rangle\mathbf{e}_z,\\
\nonumber{M}_n(\mathbf{k})&=-i\frac{e}{\hbar}\langle\mathbf{\nabla}_{\mathbf{k}}\Phi_n|\times [\epsilon_n(\mathbf{k})-H(\mathbf{k})] |\mathbf{\nabla}_{\mathbf{k}}\Phi_n\rangle\mathbf{e}_z,
\label{eqn:Berry}
\end{align}
where $\mathbf{\nabla}_{\mathbf{k}}=(\partial_{k_x},\partial_{k_y})$ and $"\times"$ denotes the two-dimensional cross product. We plot the Berry curvature and the magnetic moment of the lower conduction band in the $K_-$-valley in the lower panels of Fig.~\ref{fig:Berry} for $\Delta=150$ meV. In the $K_+$-valley both $\Omega$ and $M$ carry the opposite sign. A non-zero orbital magnetic momentum behaves like the electron spin~\cite{Xiao2010} and will hence couple linearly to a magnetic field through a Zeeman-like term $-\mathbf{M}(\mathbf{k})\cdot\mathbf{B}$.

\section{Numerical diagonalization inside the channel}
\label{sec:diag}
In the presence of a nontrivial confinement potential we diagonalize the Hamiltonian in Eq.~\ref{eqn:H} numerically in a basis of harmonic oscillator wave functions $\psi_n(x)=N_n e^{-\frac{1}{2}(\alpha x)^2}\mathcal{H}_n(\alpha x)$, where $N_n= \sqrt{\frac{\alpha}{\sqrt{\pi}2^n n!}}$ is the normalization constant and $\alpha$ is a scaling factor of unit length$^{-1}$; we choose $\alpha$ adapted to the potential $U(x)$ obtained from comparing a parabolic potential to $U(x)$. We assume free propagation of the electrons in the $y$-direction. The basis states are then of the form $\Psi_n(\mathbf{r})=e^{ik_y y }\Phi_{n}(x)$ where the $x$-dependent part is given by 
\begin{equation}\nonumber\Phi_{n,1}(x)=
\begin{pmatrix}
\psi_n(x)\\
0\\
0\\
0
\end{pmatrix},\;
\Phi_{n,2}=
\begin{pmatrix}
0\\
\psi_n(x)\\
0\\
0
\end{pmatrix},
\end{equation}
\begin{equation}
\Phi_{n,3}=
\begin{pmatrix}
0\\
0\\
\psi_n(x)\\
0
\end{pmatrix},
\Phi_{n,4}=
\begin{pmatrix}
0\\
0\\
0\\
\psi_n(x)
\end{pmatrix}.
\label{eqn:Basis}
\end{equation}

For every set of system parameters we construct the matrix
corresponding to Hamiltonian $H^{\xi}_{BLG}$ in the basis given in
Eq.~\ref{eqn:Basis} and obtain the energy spectrum by
diagonalization. Convergence is reached when the energy levels do not
change anymore upon including a higher number of basis states. In
order to include a magnetic field we do Peierls substitution in the
Hamiltonian in Eq.~\ref{eqn:H} : $\pi \rightarrow \pi -\frac{e}{c}
\mathbf{A}$. For a magnetic field perpendicular to the BLG sheet and
to preserve translational invariance in the $y$-direction we chose
Landau gauge of the form $\mathbf{A}=(0,-Bx,0).$ The basis states of
Eq.~\ref{eqn:Basis} then translate into the basis of Landau level wave
functions localized at Landau orbital $x_0=0$.

\section{Channel spectra for different parameters}
\label{sec:para}

\begin{figure*}[h!]
  \includegraphics[width=0.85\textwidth]{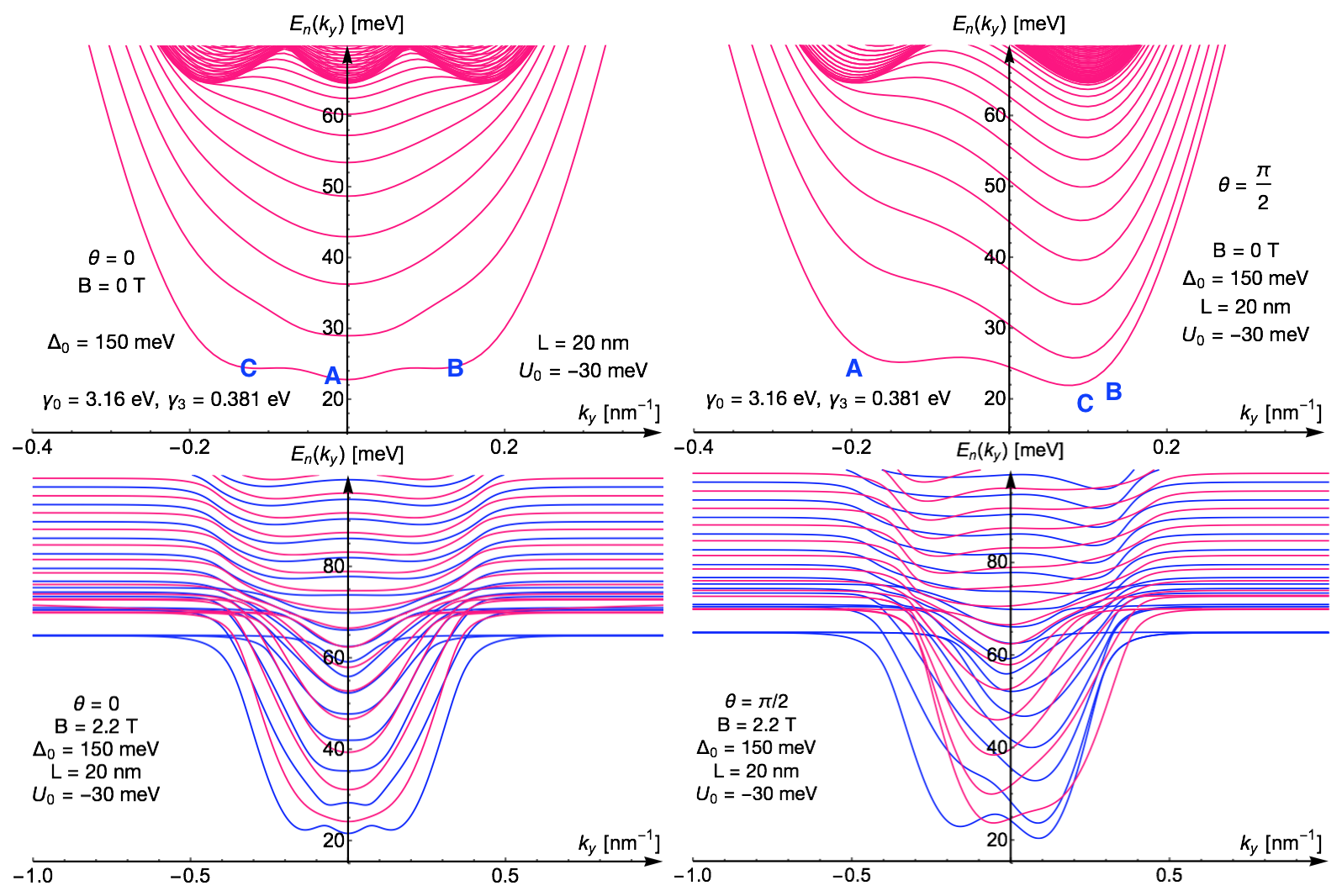}
  \caption{Conduction band spectra inside the channel for the two
    different orientations $\theta=0$ and $\theta=\frac{\pi}{2}$. Top:
    Spectra in the absence of a magnetic field. The letters $A,B$, and
    $C$ refer to the minivalleys indicated in
    Fig.~\ref{fig:Berry}. Bottom: Magnetic spectra for a magnetic
    field of $B=2.2$ T. In the presence of a non-trivial gap and a
    non-zero magnetic field valley symmetry is broken and we obtain
    two unrelated spectra for the $K_+$ valley (blue lines) and the
    $K_-$ valley (magenta lines), where the splitting between valleys
    is given by the Zeeman splitting of magnetic moment-carrying
    states in the minivalleys of gapped BLG. }
  \label{fig:SpectrHiske}
\end{figure*}

\begin{figure*}[h!]
  \includegraphics[width=0.85\textwidth]{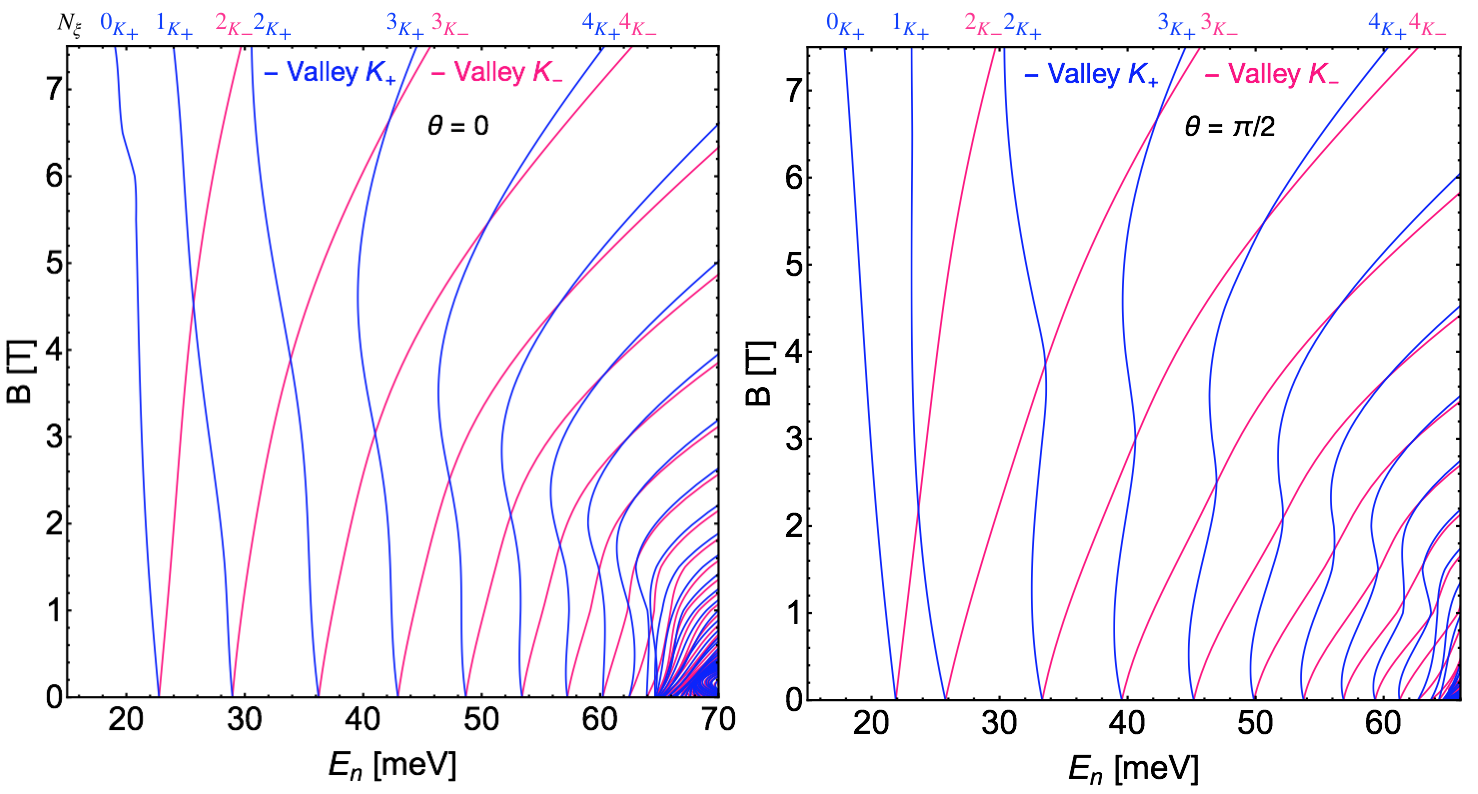}
  \caption{Conduction band edges as a function of the magnetic field strength for the same system parameters as indicated in Fig.~\ref{fig:SpectrHiske} and the two different channel orientations $\theta=0$ and $\theta=\frac{\pi}{2}$. We observe linear behaviour of the levels at low magnetic field due to the linear Zeeman coupling induced by the magnetic moment (due to the finite Berry curvature) of the states in the minivalleys of gapped BLG. At high magnetic fields, the levels evolve into the LLs of gapped BLG.   }
    \label{fig:SpectrHiskeEwB}
\end{figure*}

Due to the trigonal warping effect for non-zero $v_3$ the dispersion
is not rotationally symmetric (\textit{c.f.~}the dispersion of
homogeneous gapped BLG in Fig.~\ref{fig:Berry}) and the channel
spectra therefore depend on the orientation of the channel. We
distinguish between the two angles of orientation $\theta=0$ and
$\theta=\frac{\pi}{2}$ for which the orientation of the channel is
indicated by the blue bars in Fig.~\ref{fig:Berry}. In
Fig.~\ref{fig:SpectrHiske} we show additional channel spectra for
different system parameters and the two different angles of
orientation. Figure \ref{fig:SpectrHiskeEwB} shows the dependence of
the lower band edges as a function of the magnetic field for both
angles.  \makeatletter\onecolumngrid@push\makeatother

\clearpage
\makeatletter\onecolumngrid@pop\makeatother

\section{Details of the tight binding simulation}
\begin{figure}
	\centering
	\includegraphics[width=\columnwidth]{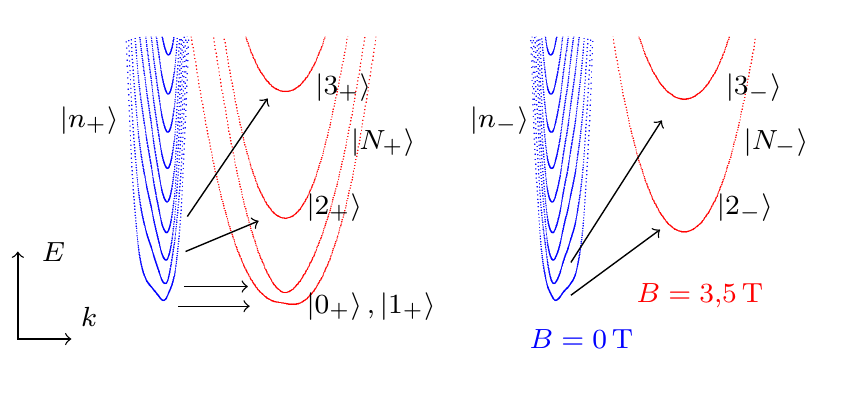}
	\caption{{{Bandstructure at $B=0~$T
			\textcolor{blue}{(blue)} featuring size quantized
			energy levels $\ket{n^S_{\xi}}$ and Landau states
			$\ket{n^L_{\xi}}$ at $B=3.5~$T
			\textcolor{red}{(red)} $\ket{n^S_{\xi}}$.  Two zero
			modes form (for $B>0$) only in the $K_+$ valley. The analytical form of
			the states is given in Eq. \ref{Lwf}}}}
	\label{f:Bandstruct}
\end{figure}

Our tight binding calculation is done for a
280~nm wide BLG nanoribbon in the
parametrisation given by Jung et al.~\cite{Jung2014}.
An additional displacement field, which we obtain from solving
the Poisson equation for the experimental setup \textsc{QPC L}
at constant channel voltage $V_{\text{ch}} = -2.5~$V, is added. 
It confines the wavefunctions to the 180~nm wide region between the gates, where the displacement field is about 50~meV.
A Berry-Mondragon type potential at the sides of the ribbon is used to eliminate edge states, restricting the simulation effectively to a width of 250~nm.
We then solve the eigenvalue problem for the Bloch state~\cite{Sanvito1999}
\begin{equation}
\kl{\mathcal H_0 + \mathrm e^{\mathrm i k \Delta x} \mathcal H_I + \mathrm e^{-\mathrm i k \Delta x} \mathcal H_I^\dagger } \chi = E \chi,
\end{equation}
for a given $k$ numerically, which has wavefunctions $\psi =
\chi \exp\kl{\mathrm i k \Delta x}$ of an infinite waveguide as
solutions. 
The bandstructure is given by the eigenvalues $E(k)$, see Fig. \ref{f:Bandstruct}.
The matrix $\mathcal H_0$ contains all on-site energies
and hopping matrix elements of a slice in $y$ direction, and
the matrix $\mathcal H_I$ contains the hopping matrix elements
between a given slice and the one to the right, separated by
$\Delta x$.  As direction of propagation we choose an angle
$\phi = \arctan(1/(2 \sqrt 3)) \approx 0.28~$rad away
from armchair direction. In armchair direction, the two cones
lie on top of each other in momentum space, making a clean
separation of $K_+$ and $K_-$ states challenging.  

\section{Effective low-energy {Hamiltonian}}
The minimal ingredients which lead to the observed crossing pattern are already present in the effective low-energy {Hamiltonian} of BLG. 
For completeness, we present a short discussion.

We consider {Bernal}-stacked BLG, with $A$ and $B$ atoms in the lower layer, and $A'$ (coupled to $B$) and $B'$ atoms in the upper layer.
Eliminating the dimer state components leads to an effective low-energy {Hamiltonian} written in terms of the wavefunction on the unpaired ($A$ and $B'$) atoms,
%
A magnetic field $\bm B = (0,0,B)$, $B > 0$, can be added with the minimal coupling prescription $\hat {\bm p} \rightarrow \hat {\bm p} - q \bm A/c$. With the confinement in $y$ direction in mind, we choose the gauge $\bm A = (-B y,0,0)$.
The {Hamiltonian} then reads
\begin{equation}
\mathcal H_{2,\xi}   =	 -\frac {1} {2m \lambda_B^2} \begin{pmatrix}
& \kl{\xi \tilde y - \mathrm i p_{\tilde y} }^2 \\
\kl{\xi \tilde y + \mathrm i p_{\tilde y} }^2    & 
\end{pmatrix}
\end{equation} 
where we define the magnetic length $\lambda_B = \sqrt{\hbar c/(e B)}$ and $\tilde y = y / \lambda_B + \lambda_B k_x$, $m = \gamma_1 / (2 v^2)$, interlayer coupling $\gamma$, Fermi velocity $v$, valley index $\xi = \pm 1$.
It has the solutions \cite{McCann2006a} for $K_+$ and $n \ge 2$
\begin{equation}
E_{n,\pm} = \pm \hbar \omega_c \sqrt{n(n-1)}, \quad  \ket{n^L_{K+}}_\pm = \frac{1}{\sqrt{2}}  \begin{pmatrix}
\phi_n \\
\pm \phi_{n-2} 
\end{pmatrix}.
\label{Lwf}
\end{equation} 
$\omega_c = 1/(m \lambda_B^2)$
and two zero-energy solutions
\begin{equation}
E = 0 \qquad  \ket{(n=0,1)^L_{K+}}_\pm =  \begin{pmatrix}
\phi_n \\
0
\end{pmatrix}. 
\end{equation}
The localization of the lowest lying Landau on a single sublattice results -- due to the absence of nearest neighbors -- in zero-energy solutions of the effective Hamiltonian.
Near $K_-$ we find the same spectrum but with the $A$ and $B'$ sublattices reversed.

\paragraph{Magnetic Field and displacement field}
A displacement field $\sigma^z\Delta /2$ acts as an effective mass term and lifts the $K_\xi$ degeneracy\cite{McCann2006a},
\begin{equation}
E_{n,\pm} = \pm \hbar \omega_c \sqrt{n(n-1)} - \xi \frac{\Delta}{2} \frac{\hbar \omega_c}{\gamma_1}, \qquad n \ge 2.
\end{equation} 
The two zero modes are shifted above ($K_+$) or below ($K_-$) the gap\cite{McCann2006a}, 
\begin{equation}
E_0 = \xi \frac{\Delta}{2}, \qquad E_1  = \xi \frac{\Delta}{2} - \xi\frac{\Delta \omega_c}{\gamma_1}.
\end{equation}
The states which are localized on one sublattice and thus localized on one of the layers are trivially affected by the electrostatic potential: The zeroth Landau levels where $\xi B$ is positive gets shifted above the gap, and the level where  $\xi B$ is negative is shifted below the gap.

\section{Monolayer Graphene}\label{app:mono}
\begin{figure}
		\centering
		\includegraphics[width=\columnwidth]{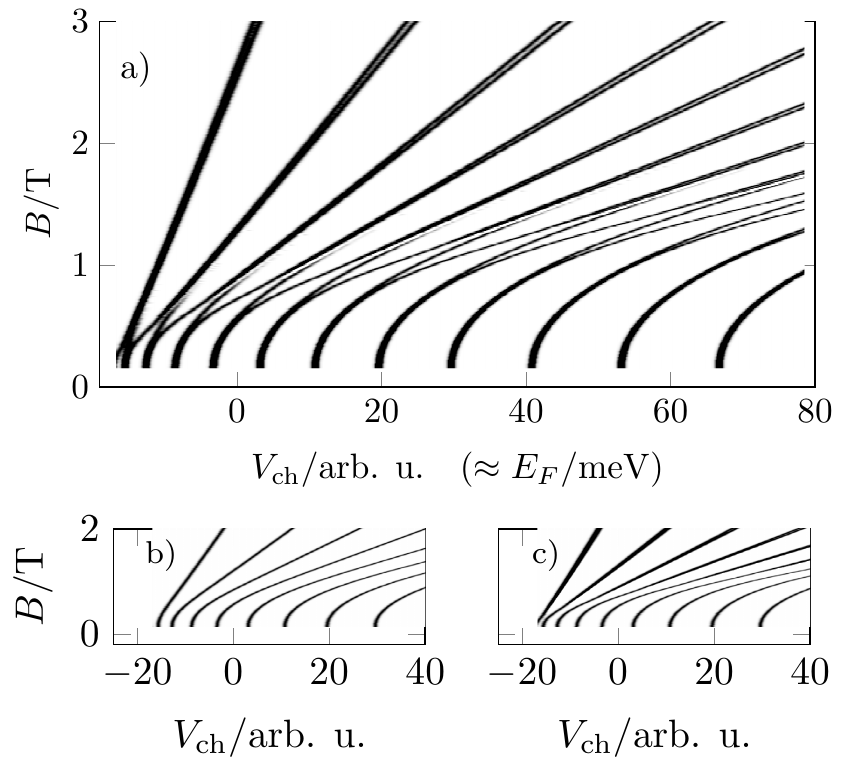}
	\caption{
		{{
		a) Differential conductance $\mathrm d G / \mathrm d E$ in a 280~nm wide monolayer graphene ribbon, including a thermal smoothening of 1.7~K. 
		A potential confines the states to a region of 180~nm.
		A Berry-Mondragon type potential at the sides of the ribbon is used to eliminate edge states.
		b) and c) show separately the contributions from the two valleys $K_+$ and $K_-$ at low energies.} 
	}}
	\label{f:MonolayerGraphene}
\end{figure}

A similar effect can be observed in monolayer graphene, see Fig. \ref{f:MonolayerGraphene}.

The only difference is that due to the existence of a single zero mode instead of two in the quantum Hall regime, the evolution now follows the pattern
$\ket{(n-1)^S_{-}} \rightarrow \ket{n^L_{-}}$ and $\ket{n^S_{+}}
\rightarrow \ket{n^L_{+}}$ for positive $B$, and with the roles of $K_+$ and $K_-$ reversed for negative $B$.

\clearpage
\section{Transconductance of QPC S}
\label{sec:qpcs}
\begin{figure}
\centering
\includegraphics[width=0.5\textwidth]{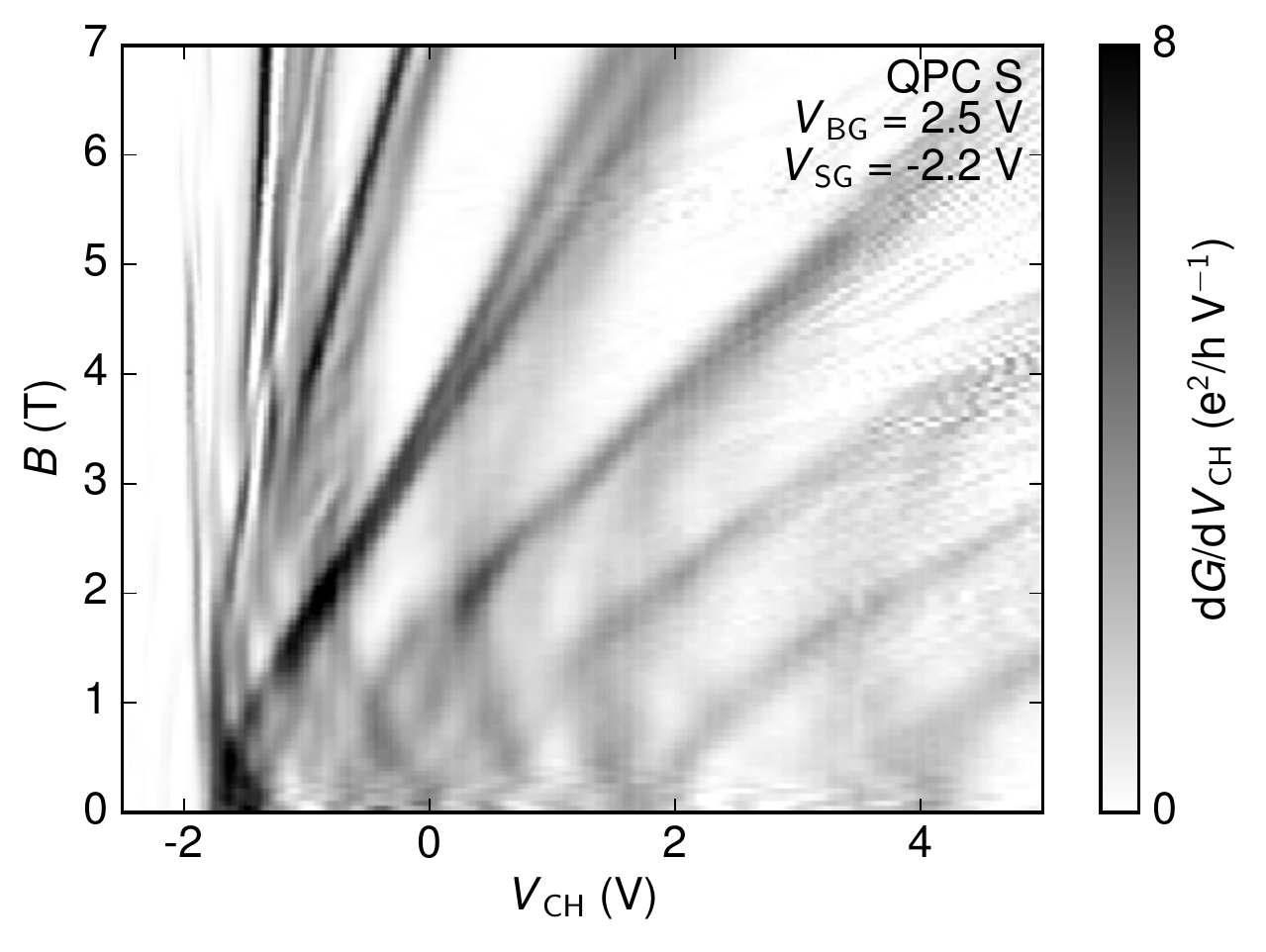}
\caption{Transconductance of QPC S as a function of channel gate voltage and magnetic field. Several mode crossings can be observed. The spacing between the modes is larger than for QPC M and QPC L, as expected for a narrower channel.}
\label{fig:QPCS}
\end{figure}

The transconductance of QPC S as a function of channel gate voltage and magnetic field is shown in Fig.~\ref{fig:QPCS}. It shows a pattern of mode crossings similar to the patterns of QPC M and QPC L (see Fig.~\ref{fig:fans}b,c), with a larger mode spacing due to the narrower confinement potential.

\section{Transconductance in a lower displacement field}
\label{sec:displ}
\begin{figure}
\centering
\includegraphics[width=0.5\textwidth]{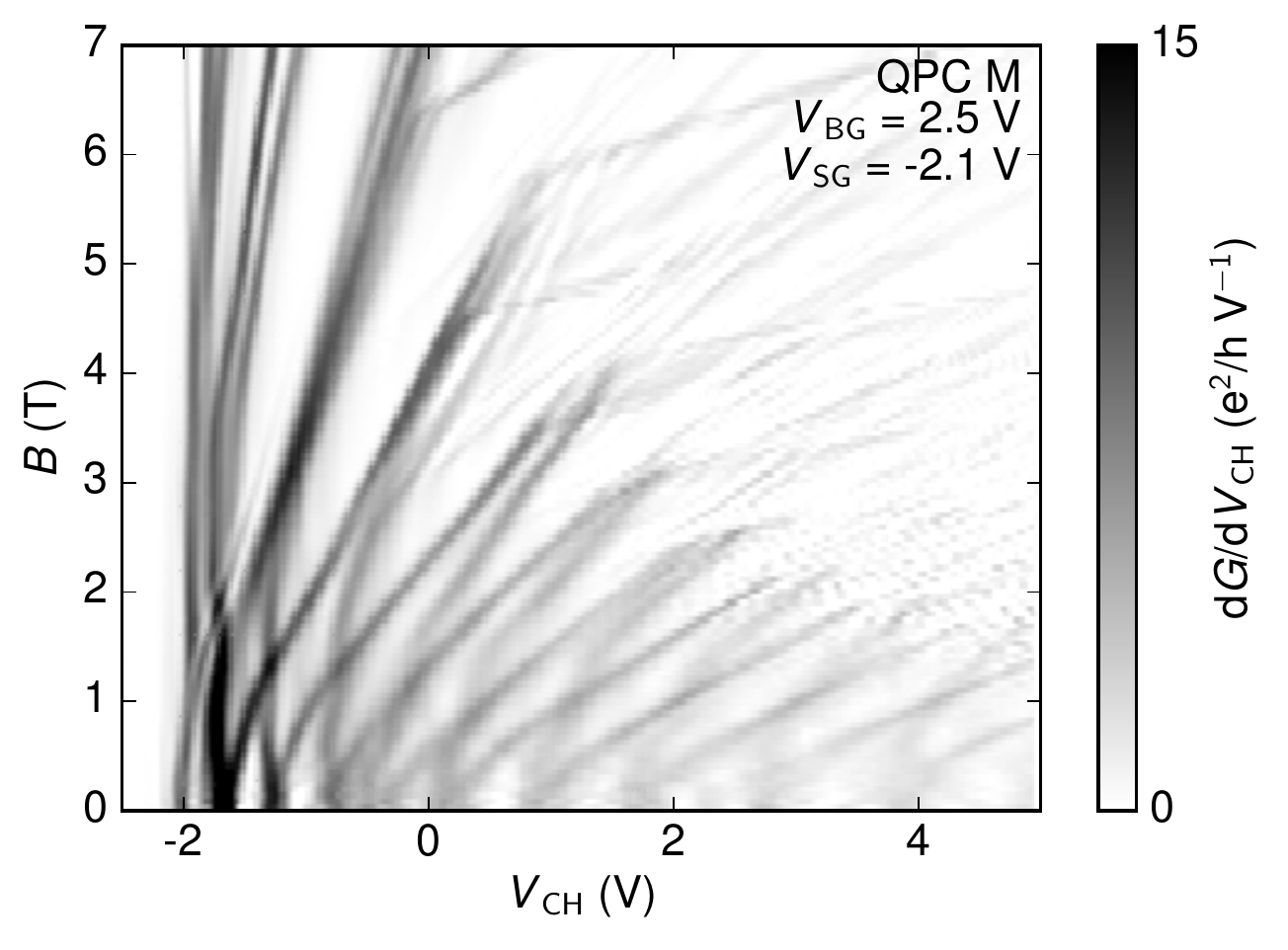}
\caption{Transconductance of QPC M as a function of channel gate voltage and magnetic field for $V_\mathrm{BG} = 2.5~V$. The observed pattern of mode crossings is similar to the pattern in Fig.~\ref{fig:fans}b.}
\label{fig:lowD}
\end{figure}

Figure~\ref{fig:lowD} shows the transconductance of QPC M as a function of channel gate voltage and magnetic field for $V_\mathrm{BG} = 2.5~V$ and $V_\mathrm{BG} = -2.2~V$. Although this corresponds to a smaller displacement field in the channel than in Fig.~\ref{fig:fans}b, a similar pattern of mode crossings is observed. No modes can be observed in the top right corner of Fig.~\ref{fig:lowD}, because in this regime the filling factor in the bulk is smaller than the filling factor in the channel.
\section{Transconductance in the $p$-type regime}
\label{sec:ptype}
\begin{figure}
\centering
\includegraphics[width=0.5\textwidth]{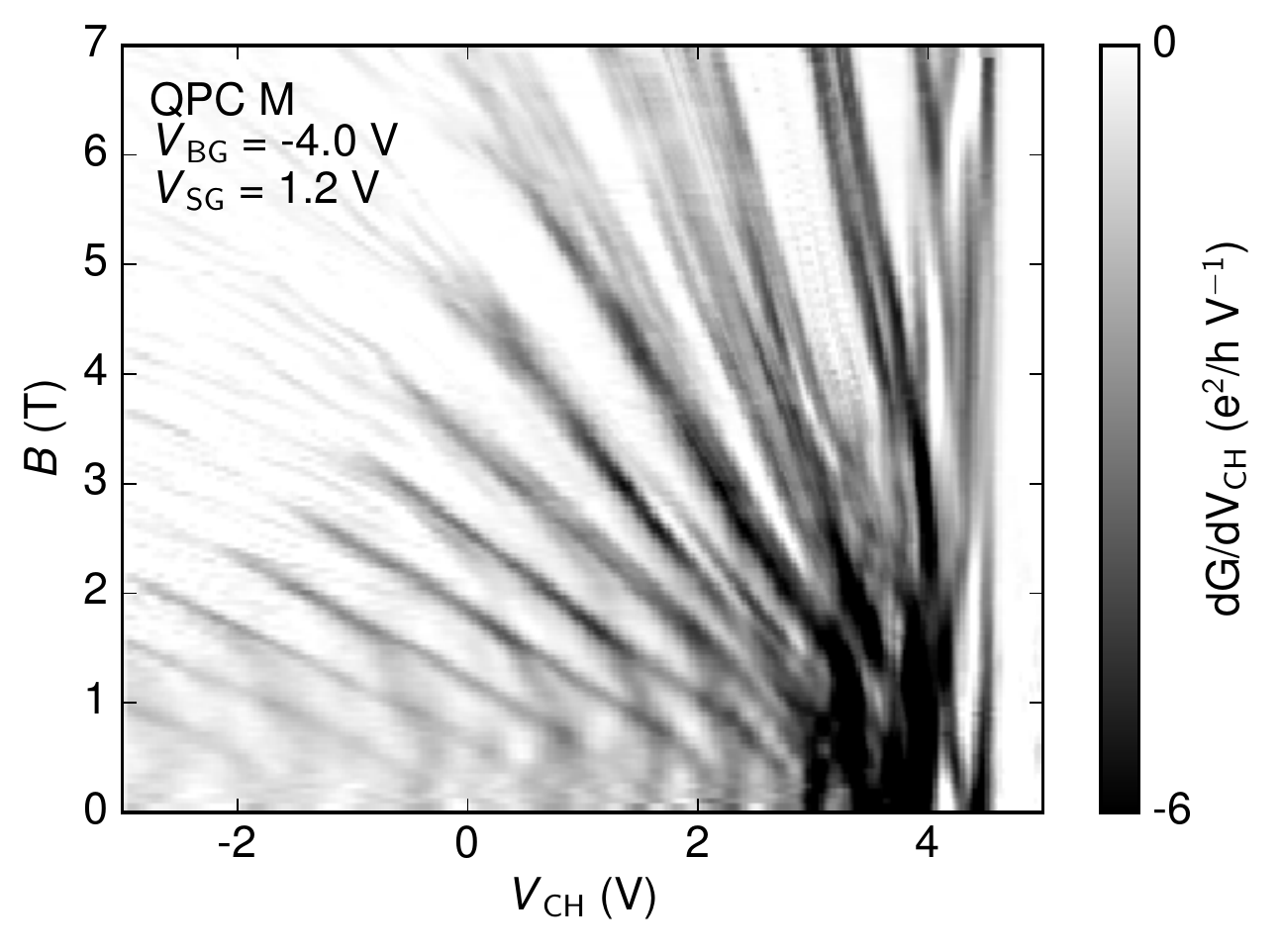}
\caption{Transconductance of QPC M as a function of channel gate voltage and magnetic field for a $p$-type channel. For the higher modes (at $V_\mathrm{CH} < 2~$V) the pattern of mode crossings looks similar to that observed in Fig.~\ref{fig:fans}b.}
\label{fig:holes}
\end{figure}

The pattern of mode crossing can also be observed for a $p$-type channel, as shown for QPC M in Fig.~\ref{fig:holes}. 

\section{Finite bias diamonds}
\label{sec:fb}

\begin{figure}
\centering
\includegraphics[width=0.5\textwidth]{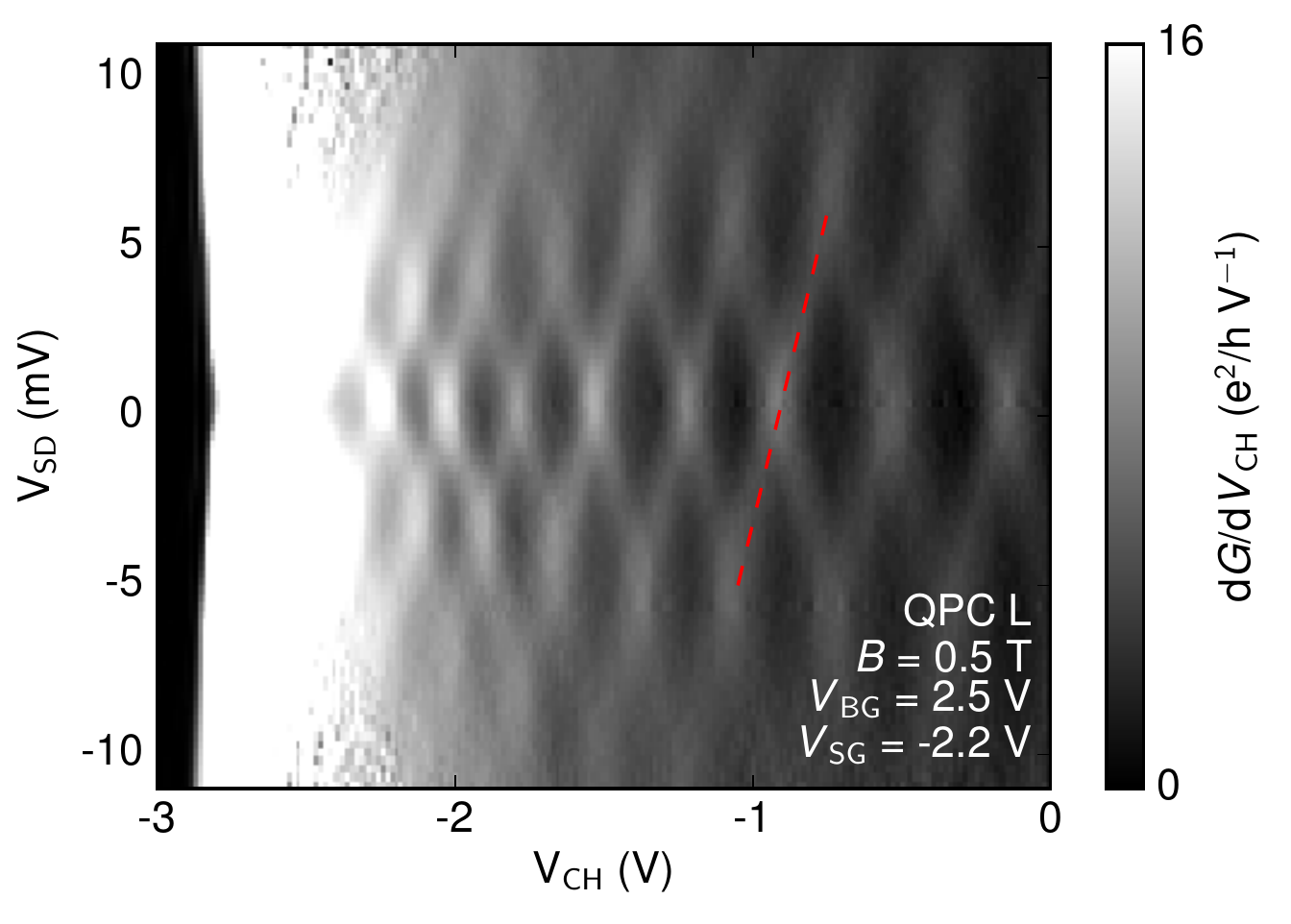}
\caption{Finite bias spectroscopy of QPC L. A typical diamond pattern is observed. From the slope of finite bias diamond boundaries, the lever arm of the channel gate can be determined.}
\label{fig:diamonds}
\end{figure}

Figure \ref{fig:diamonds} shows the finite bias measurements of the modes of QPC L, showing a mode spacing on the order of 3 meV at $B = 0.5~$T. From the slope of the diamond boundaries (see red dashed line in Fig.~\ref{fig:diamonds}), a lever arm $\alpha$ for conversion from channel gate voltage to energy can be extracted ($E = \alpha e V_\mathrm{CH}$). For QPC L we find $\alpha = 18 \times 10^{-3}$ and for QPC M we find $\alpha = 14 \times 10^{-3}$. The smaller lever arm for the narrower QPC is due to more significant screening of the channel gate voltage for this QPC.

\end{document}